\documentstyle[aps,prl,epsfig,multicol,amssymb]{revtex}
\begin{document}
\draft

 \title{Generalized effective depletion potentials} 
\author{A.A. Louis$^{a}$ and  R.Roth$^{b}$ }
\address{$^a$Department of Chemistry, Lensfield Rd, Cambridge CB2 1EW, UK}
\address{$^b$H.H. Wills Physics Laboratory, University of Bristol, Bristol
BS8 1TL, UK}
 \date{\today} \maketitle
\begin{abstract}
\noindent
We propose that the behavior of asymmetric binary fluid mixtures with
a large class of attractive or  repulsive interparticle interactions
can be understood by mapping onto effective non-additive hard-sphere
models.  The latter are best analyzed in terms of their underlying
depletion potentials which can be {\em exactly} scaled onto known
additive ones. By tuning the non-additivity, a wide variety of
attractive or repulsive generalized depletion potential shapes and
associated phase behavior can be ``engineered'', leading, for example,
to two ways to stabilize colloidal suspensions by adding smaller
particles.

\end{abstract} \vglue-0.3cm

\begin{multicols}{2}
Colloidal suspensions are complex mixtures of mesoscopic solute
particles and smaller solvent particles.  Examples include a wide
variety of organic or inorganic solutes, ranging from proteins to
micelles to polymeric composites to ceramic materials etc...,
suspended in polar or non-polar solvents. Varying the interactions
between the constituent particles results in a broad range of
equilibrium and non-equilibrium fluid behavior.  This tunability has
led to the wide spread industrial and biological applications of
colloidal suspensions\cite{Russ89}. Some more recent developments
include the design of complex self-assembled materials such as
photonic band-gap crystals by use of templates\cite{Dins96}, and new
experimental advances that allow colloidal interactions to be directly
measured with a greatly increased accuracy\cite{Isra92}.

In all the applications mentioned above, the design of a colloidal
fluids with certain desired properties requires control over the inter
particle interactions.  These interactions are typically effective,
that is to say they are a combination of direct interactions (such as
Coulomb forces) with indirect interactions mediated through the
solvent and the other solute particles\cite{Bell00,Liko01,Loui01a}.
One of the best known is the indirect depletion interaction, where one
set of (typically smaller solute or solvent) particles induces an
interaction between another set of particles.  Depletion potentials
were first calculated for mixtures of polymers and
colloids\cite{Asak54} and, with the advent of new experimental and
theoretical techniques, they have been the subject of intense recent
interest\cite{Bibe91,Lekk92,Dijk98,Dijk99,Buho98,Roth00a}

Theoretical work has often focussed on the binary hard-sphere (HS)
model, for which a depletion induced phase separation for
size-ratios's $q = \sigma_{ss}/\sigma_{bb} < 0.2$ was
suggested\cite{Bibe91} ($\sigma_{\alpha\alpha}$ is the diameter of the
big ({\em b}) or small ({\em s}) particles). A key advance was made by
Dijkstra {\em et al.\ }\cite{Dijk98}, who used an effective
one-component depletion potential picture to show that the fluid-fluid
phase-separation found with a two-component integral equation
technique by Biben and Hansen\cite{Bibe91} was metastable w.r.t.\ a
fluid-solid phase-transition.  More generally, their approach added to
the growing consensus that a carefully derived effective potential is
a powerful tool for analyzing the  behavior of an asymmetric binary
mixture, at least 

\noindent 
for size-ratios $q \leq 0.3$ where many-body
interactions are not thought to be important (see e.g.\
\cite{Bell00,Liko01,Loui01a} for some recent reviews).  The key step
in all these approaches is integrating out the smaller component of a
binary mixture to leave a new one-component system with an effective
interaction between the big particles.

 Most theories of depletion have considered only hard-core interactions
leading to purely entropic depletion potentials. Their range varies
with $\sigma_{ss}$ while increasing the small particle density
$\rho_s$ or packing fraction $\eta_s = \pi \rho_s \sigma_{ss}^3/6$
increases the depth of the (always) attractive well at contact, and
possibly adds enhanced oscillations at larger separations
$r$\cite{Asak54,Lekk92,Roth00a}.

There have been a number of recent attempts to go beyond purely
entropic depletion by including extra interactions between the
particles of a binary HS
mixture\cite{Walz94,Amok98,Malh99,Clem00,Mend00}.  Of course many
different kinds of extra interactions can be added, leading to a
seemingly enormous increase in complexity.  However, in this letter we
propose that the effect of a wide variety of these extra interactions
on depletion potentials can be understood by a simple mapping onto a
non-additive HS mixture model, for which the depletion potentials can
be calculated by a second {\em exact} mapping or scaling onto those of
an additive system.

Since the phase behavior of many binary fluids can be well understood
on the basis of these depletion
potentials\cite{Bell00,Liko01,Loui01a}, this implies that our (double) mapping
can be
used to analyze a wide variety of interacting asymmetric binary
mixtures.  These ideas can also be turned around, leading to the
possibility of explicitly engineering a wide variety of {\em
generalized depletion potential} shapes, including potentials that are
repulsive at contact, by not only varying the usual parameters
$\rho_s$ and $q$, but also by tuning the interparticle interactions to
vary the non-additivity.

\begin{figure}
\begin{center}
\epsfig{figure=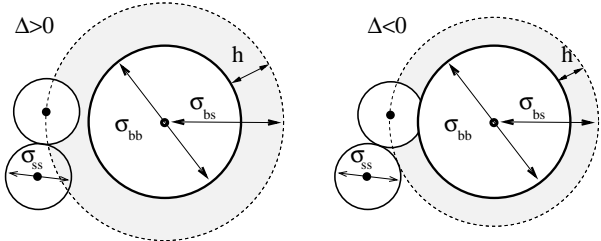,width=3.2cm,angle=-90}
\begin{minipage}{8cm}
\caption{\label{Fig1}Centers of the small-particles of diameter
$\sigma_{ss}$ are excluded from the shaded depletion layer of width
$h$ around each big particle of diameter $\sigma_{bb}$.  If $h <
\frac12 \sigma_{ss}$ then $\Delta < 0$ (negative non-additivity).  If
$h > \frac12 \sigma_{ss}$, then $\Delta > 0 $ (positive
non-additivity) }
\end{minipage}
\end{center}
\end{figure}

Non-additive binary HS models are defined by specifying the
cross-diameter\cite{Hans86}
\begin{equation}\label{eq1}
\sigma_{bs} = \frac{1}{2} \left( \sigma_{ss} + \sigma_{bb} \right)\,
\left( 1 + \Delta \right).
\end{equation}
When $\Delta =0$, the model follows the Lorentz mixing rule, and is
traditionally called {\em additive} (not to be confused with pairwise
additivity of potentials), that is the cross-diameter is simply the
sum of the two radii, exactly what one would expect on purely
geometric grounds. If $\Delta > 0$ or $\Delta < 0$ the system shows
{\em positive} or {\em negative non-additivity} respectively.  As
shown in Fig.~\ref{Fig1}, each big particle excludes a volume $v_b =
\pi \sigma_{bs}^3/6$ from the centers of the smaller particles.  When
the depletion layers of the two big particles (width defined as $h =
\sigma_{bs}-\sigma_{bb}/2 =\frac12(\sigma_{ss} + \Delta (\sigma_{ss} +
\sigma_{bb}))$) begin to overlap, then the small particles can gain
free volume $v_\Delta$, leading to a depletion interaction. 
\vglue - 0.7cm
\vglue - 0.7cm

To calculate these potentials we first note that the depletion
potential $\beta V_{eff}(r)$ depends only on the big-small and
small-small interactions ($\beta V_{bs}(r)$ and $\beta V_{ss}(r)$
respectively), but not on any direct big-big interaction $\beta
V_{bb}(r)$, which can simply be added to the depletion
potential\cite{Loui01a,Roth00a}. For non-additive systems at fixed
$\rho_s$ this means that the depletion potential 
is determined by $\sigma_{bs}$ and $\sigma_{ss}$, and is equivalent to
an additive one with the same parameters! Only the cutoff due to
$\sigma_{bb}$ is different.  For example, if one has an expression for
the additive potential in terms of $\rho_b$, $q$ and the scaled
distance $p=r/\sigma_{bb}$, then the a potential with $\Delta \neq 0$
is given by: $\beta V_{eff}(\rho_s,q,\Delta,p) = \beta
V_{eff}(\rho_s,q',\Delta=0,(q'/q)p )$ where $q'=
\sigma_{ss}/(2\sigma_{bs} - \sigma_{ss}/2)$.  Details of this (perhaps
surprising) exact mapping by scaling to additive potentials will be
given elsewhere\cite{Roth01b}.
%
 The depletion potentials for additive systems can be calculated to
quantitative accuracy by a density functional theory (DFT)
technique\cite{Roth00a}, and we checked that the scaling procedure
above exactly reproduces recent DFT calculations of non-additive
depletion potentials\cite{Roth01a}.

\begin{figure}
\begin{center}
\epsfig{figure=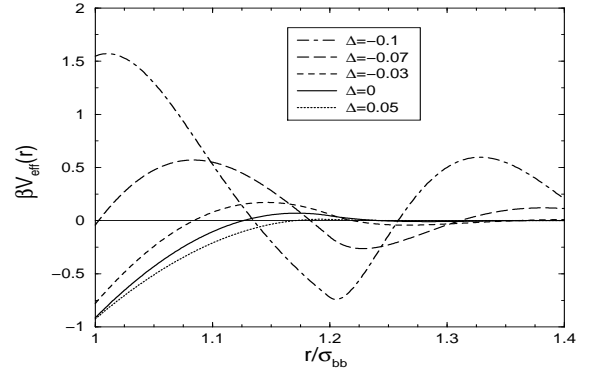,width=7.5cm,height=5cm} 
\begin{minipage}{8cm}
\caption{\label{Fig2} Depletion potentials when parameters are changed
according to {\bf case (A)}, i.e.\ varying $\sigma_{ss}$ but keeping
$h$ constant.  Here $q=0.2$, and $4 \pi \rho_s h^3/3= 0.1$. The
potentials are calculated by an exact scaling from known additive DFT
results\protect\cite{Roth00a}.}
\end{minipage}
\end{center}
\end{figure}

For a fixed number density $\rho_{s}$, non-additivity can be
introduced in two ways:

{\bf Case (A)} Fix the depletion layer width $h$, (or equivalently
$\sigma_{bs}$), and vary $\Delta$ by changing the small particle
diameter $\sigma_{ss}$.  The effect on depletion pair-potentials
$\beta V_{eff}(r)$ is shown in Fig.~\ref{Fig2}.  For increasing
positive non-additivity the correlation induced maximum decreases and
the potential tends towards the (ideal) Asakura-Oosawa
(AO)\cite{Asak54} limit; the contact value remains relatively
constant, as was found earlier\cite{Roth01a}.  In contrast, for
increasing negative non-additivity the contact value increases
markedly, leading to the possibility of strongly repulsive
interactions.  A naive application of the simplest depletion picture
where $\beta V_{eff}(r=\sigma_{bb}) = -\Pi_s v_\Delta$ would give the
opposite effect, since decreasing $\Delta$ by increasing $\sigma_{ss}$
increases the packing fraction $\eta_s = \pi \rho_s \sigma_{ss}^3/6$
and therefore the small particle osmotic pressure $\Pi_s$, while keeping
$v_\Delta$ unchanged, seemingly leading to a more attractive effective
potential.  However, a more careful analysis reveals that increasing
$\eta_s$ leads to well-developed solvation shells around a single big
particle.  When two big particles approach, the overlap of the
solvation shells leads to the repulsive interactions, as well as
larger oscillations in the pair potentials.  We found that the
amplitude of this repulsion becomes larger for smaller size-ratios,
with values possible of many times $k_BT$.

\begin{figure}
\begin{center}
\epsfig{figure=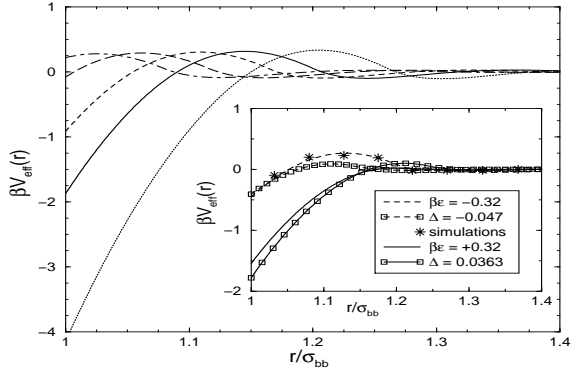,width=7.5cm,height=5cm} 
\begin{minipage}{8cm}
\caption{\label{Fig3}Depletion potentials when parameters are changed
according to {\bf Case (B)}, i.e.\ varying $h$ but keeping
$\sigma_{ss}$ constant.  Here $q=0.2$, $\eta_s = 0.2$ and line-styles
denote the same $\Delta$s as in Fig.~\protect\ref{Fig2}.  Inset:
Depletion potentials when $\beta V_{bs}(r) = \epsilon
\exp[-\kappa(r-\sigma_{bs})]/r$ is added to a binary HS mixture are
compared to the non-additive HS case with $\Delta$ calculated by our
simple mapping.  Here $\eta_s=0.116$, $q=0.2$, $\kappa \sigma_{ss} =
4$ is fixed and $\beta \epsilon$ is varied. The included simulation
data for $\beta \epsilon=-0.32$\protect\cite{Malh99} helps confirm
the accuracy of our new direct DFT approach.}
\end{minipage}
\end{center}
\end{figure}

 {\bf Case (B)} Fix the small-particle hard-core diameter
$\sigma_{ss}$, and vary $\Delta$ by changing $h$ (or equivalently
$\sigma_{bs}$).  The dominant effect on depletion pair-potentials is to 
shift them along $r$ as shown in
Fig.~\ref{Fig3}.  In this case both positive and negative
non-additivity change the well-depth at contact significantly because
changing $h$ changes the amount of volume doubly excluded when two big
particles approach. This can be illustrated at the simple AO level
where the potential at contact is given by
\begin{equation}\label{eq2}
\beta V_{AO}(r=\sigma_{bb}) = -\rho_s \frac{\pi}{4} \left( \sigma_{bb}(2
h)^2 + \frac{2}{3} (2 h)^3 \right).
\end{equation}
 On the other
hand, the correlation induced maximum remains roughly the same since
$\eta_s$ is constant, leading to similar solvation layers of the small
particles around a big particle. 
 \vglue - 0.4cm
\vglue - 0.6cm


With the insight gained from analyzing non-additive depletion
potentials, we next pose the question: what happens to $\beta
V_{eff}(r)$ for a given binary HS mixture when more general attractive
or repulsive interactions $\beta V_{ss}(r)$ or $V_{bs}(r)$ are added?
Inspired by some well-established ideas from the theory of simple
liquids\cite{Hans86}, we map onto effective HS diameters as follows:
$\sigma_{\alpha \beta} = \sigma_{\alpha \beta}^0 + \int (\exp[-\beta
V_{\alpha\beta}(r)]-1) dr$, a procedure similar to the well known
Barker-Henderson approach\cite{Bark67}. Here $\sigma_{\alpha \beta}^0$
denotes the original effective HS diameters without the extra
interaction.  In this way the additional interactions can be mapped
onto an effective non-additive HS model as follows:

\noindent{\bf (i)}\,\,\, repulsive $\,\,\beta V_{ss}(r)$ :
$\sigma_{bs}=\sigma_{bs}^0$; $\sigma_{ss}> \sigma_{ss}^0$; $\Delta <
0$

\noindent{\bf (ii)}  attractive $\beta V_{ss}(r)$ :
$\sigma_{bs}=\sigma_{bs}^0$; $\sigma_{ss}<\sigma_{ss}^0$; $\Delta > 0$

\noindent{\bf (iii)} repulsive $\,\,\beta V_{bs}(r)$ :
$\sigma_{bs}>\sigma_{bs}^0$; $\sigma_{ss}=\sigma_{ss}^0$; $\Delta > 0$

\noindent{\bf (iv)} attractive $\beta V_{bs}(r)$ :
$\sigma_{bs}<\sigma_{bs}^0$; $\sigma_{ss}=\sigma_{ss}^0$; $\Delta < 0$

\vglue - 0.4cm
\vglue - 0.55cm

 The depletion potentials for {\bf (i)} ($\Delta < 0$) and {\bf (ii)}
($\Delta > 0$) change according to {\bf case (A)}, which is depicted
in Fig.~\ref{Fig2}, while the depletion potentials for {\bf
(iii)}($\Delta > 0$) and {\bf (iv)} ($\Delta < 0$) change according to
{\bf case (B)}, as depicted in Fig.~\ref{Fig3}.  This picture agrees
qualitatively with calculations of other authors of depletion
potentials for non-HS systems. Examples of the pathways above
include: {\bf (i)} Fig.~4 of\cite{Mend00}; {\bf (ii)} Fig.~8
of\cite{Amok98}, Fig.~4 of\cite{Bell00}; {\bf (iii)} Fig.~5
of\cite{Walz94}; Fig.~3 of\cite{Amok98}; (See also \cite{Mond95} for a
recent experimental and \cite{Piec00} for a recent theoretical mapping
of this type of interaction onto an AO potential).  {\bf (iv)} Fig.~4
of \cite{Amok98}, Fig.~4 of \cite{Malh99}, Fig.~7 of \cite{Clem00},
and Fig.~4 of\cite{Bell00}.  Keep in mind, however, that these
calculations were done with a number of approximate techniques which
may not always give quantitatively reliable results, especially for
contact values\cite{Loui01a}.  Even so, our mapping scheme to
non-additivity clearly qualitatively rationalizes the dominant changes
in depletion potentials caused by changing a variety of inter-particle
interactions.

   It would be interesting to make this qualitative correspondence
more quantitative.  We were able to extend the quantitatively reliable
DFT method described in \cite{Roth00a} to systems with an arbitrary
$V_{bs}(r)$ potential (details will be published elsewhere).  In the
inset of Fig~\ref{Fig3} we compare these to non-additive HS potentials
with $\Delta$ determined by our aforementioned mapping procedure.
This gives a reasonably good representation of the well depth, but
does slightly less well for the repulsive barrier.  With our very
simple mapping procedure, we expect that the quantitative agreement
will deteriorate for very strongly attractive $V_{bs}(r)$ or
$V_{ss}(r)$, but the qualitative picture should remain the same.

\begin{figure}
\begin{center}
\epsfig{figure=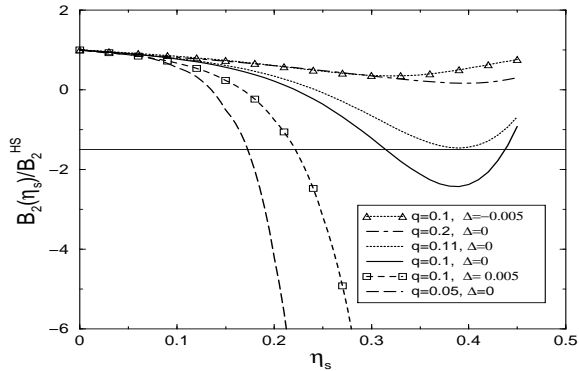,width=7.5cm,height=5cm} 
\begin{minipage}{8cm}
\caption{\label{Fig4}
Reduced second virial coefficient of the big particles $B_2/B_2^{HS}$,
plotted v.s. the packing fraction of the small particles $\eta_s$. 
Fluid-fluid phase-separation is expected when
$B_2/B_2^{HS}<-1.5$\protect\cite{Vlie00} (horizontal line).
}
\end{minipage}
\end{center}
\end{figure}

From the above it is clear that non-additivity has a profound effect
on depletion potentials, implying that this should also be reflected
in  phase-behavior.
%
For more symmetric mixtures the effect of interparticle interactions
on phase-stability has traditionally been understood in terms of
conformal-solution theory\cite{Hans86}.  For asymmetric mixtures, our
depletion potentials help generalize these ideas. For example,
Vliegenthart and Lekkerkerker\cite{Vlie00} have recently shown that
the fluid-fluid critical point of many one-component fluids occurs
when the reduced second virial-coefficient $B_2/B_2^{HS} \approx
-1.5$.  We checked that this works well for the depletion potential
simulations of Dijkstra {\em et al.\ }\cite{Dijk98,Dijk99}, suggesting
that this surprisingly accurate criterion can also be used to predict
the effect of non-additivity on fluid-fluid phase-separation.  In
Fig~\ref{Fig4} we plot the effect of the $\eta_s$ on the second virial
coefficients calculated from depletion potentials. Firstly, for the
additive case we find that $B_2/B_2^{HS} < -1.5$ only for size-ratios
$q \lesssim 0.11$ (For $q=0.1$ there might be an upper critical
point!).  But even without attributing quantitative accuracy to the
Vliegenthart Lekkerkerker criterion, the upper limit of $q$ that
allows fluid-fluid phase-separation is clearly bounded by $0.1 \leq q
\leq 0.2$, since $B_2/B_2^{HS}$ remains positive for any $\eta_s$ if
$q > 0.2$, while it goes well below $-1.5$ for $q < 0.1$.  This
finding helps in understanding earlier results obtained with
(approximate) 2-component integral equation studies\cite{Bibe91} as
well as some direct simulations\cite{Buho98}, lending support to our
argument that the underlying depletion potentials from which the $B_2$
are derived are a key to understanding the full phase-behavior of the
asymmetric two-component systems.

 Next we turn to the effect of non-additivity on the fluid-fluid
phase-separation. Fig.~\ref{Fig4} shows that for $q=0.1$ a very small
non-additivity, of the order of a $5\%$ change in $\sigma_{ss}$ or a
$0.5 \%$ change in $\sigma_{bs}$, is enough to dramatically change the
behavior of $B_2/B_2^{HS}$. For other size-ratios we find similar
effects.  For example if $\Delta = q/20$ we find that (metastable)
fluid-fluid phase-separation can occur for size-ratios up to $q=0.4$,
while if $\Delta = -q/20$, it will only occur for size-ratios $q
\lesssim 0.05$. Clearly, even a very small negative non-additivity
strongly suppresses phase-separation, while positive non-additivity
strongly enhances it. This is consistent with and helps rationalize
some earlier 2-component studies\cite{Bibe97,Dijk98b,Loui00a}.

Binary mixtures may also undergo fluid-solid phase-separation which,
for example, is the thermodynamically stable phase-transition in
additive HS mixtures\cite{Dijk98}.  Recently, one of
us\cite{Loui00a,Loui01a} has shown that for short range potentials the
fluid-solid transition shifts to low values of $\eta_b = \pi \rho_b
\sigma_{bb}^3/6$ when the potential well-depth at contact is near
$\beta V_{eff}(r=\sigma_{bb}) \approx - 2.5$; this effect is largely
independent of other details such as the range or oscillations of the
potential.  This suggests that introducing any non-additivity
according to {\bf case (B)} will strongly affect the fluid-solid
behavior. Similarly adding negative non-additivity according to {\bf
case (A)} will suppress fluid-solid phase-separation, but positive
non-additivity will not change the fluid-solid phase-boundaries much,
which is confirmed by comparing the additive ($\Delta=0$) HS to pure
AO ($q=\Delta$) simulations of Dijkstra {\em et
al.}\cite{Dijk98,Dijk99}.  \vglue - 0.3cm
\vglue - 0.6cm

The stabilization of colloidal suspensions is critical to many
industrial and experimental applications\cite{Russ89}.  The arguments
above for both fluid-fluid and fluid-solid phase-separation suggest
that the addition of smaller particles may provide such a
stabilization mechanism against demixing for colloidal suspensions as
long as $\Delta < 0$.  This can be achieved by pathway {\bf (i)},
adding a repulsive $V_{ss}(r)$, or by pathway {\bf (iv)}, adding an
attractive $V_{bs}(r)$.

In conclusion then, we combined a new approximate and a new exact
mapping to show that the non-additive HS mixture model provides an
intuitive and general organizing framework within which to understand
the effective depletion potentials induced by a large class of
interactions $V_{bs}(r)$ or $V_{ss}(r)$.  These generalized depletion
potentials can be crafted into many different shapes, and provide
access to the phase-behavior of interacting asymmetric binary
mixtures.  Clearly much more can be done by both theories and
experiments to exploit the flexibility of these potentials and to
``engineer'' desired phase-behavior in colloidal suspensions. We hope
this letter has shown some promising new directions in which to
embark.

AAL acknowledges support from the Isaac Newton Trust, Cambridge,
RR acknowledges support from the EPSRC under grant No.\ GR/L89013.
We thank R. Evans and J.P. Hansen for helpful discussions. 
\vglue -0.5cm

\end{multicols}

\end{document}